\documentclass[fleqn,10pt]{IEEEtran}
\pdfoutput=1
\usepackage{amsmath, amsthm, amsfonts}
\usepackage{graphicx}
\usepackage{epstopdf}
\title{An Exploratory Method for Smooth/Transient Decomposition}

\newcommand{\R}{\mathbb{R}}

\theoremstyle{definition}
\newtheorem{defn}{Definition}

\newtheorem{prop}{Proposition}

\usepackage{algorithm}
\usepackage[noend]{algpseudocode}

\author{\.Ilker Bayram
\\  \thanks{\.{I}. Bayram is with Analog Devices Inc., Analog Garage, Boston, MA, USA. E-mail : ibayram@ieee.org.
}
}

\date{}
\begin{document}

\maketitle
\begin{abstract}
We consider a separation problem where the observation consists of the sum of a high amplitude smooth signal and a low amplitude transient signal. We  propose a method for decomposition that relies on solving instances of a `constrained filtering problem', which is posed as a convex minimization problem. We provide a fast algorithm for solving the minimization problem, and demonstrate the potential of the scheme for a vital signs monitoring experiment using radar.
\end{abstract}
\section{Introduction}
This letter develops an exploratory signal analysis tool for accurately estimating the components of a composite signal, where precise models are not available. We specifically consider a problem where a small amplitude transient signal is mixed with a high amplitude smooth signal. The application motivating this setup is radar-based vital signs monitoring. In that application, radar picks up a 1D motion signal from a subject's chest. This signal is thought to comprise of respiration and heart activity components. Respiration is slower, but significantly higher in amplitude than the heart activity component. The heart activity component is not easy to model because its shape depends on the antenna beam pattern, position of the radar relative to the subject, radar operating frequency (which in turn determines the amount of radiation penetrating the body, if any) \cite{wil17p81}. An instance of such a signal is shown in Fig.~\ref{fig:radar}a. The main challenge is to extract the heart activity component from this signal.

\subsection*{Relevant Approaches}
Arguably the simplest method used in practice for this problem is linear-time invariant (LTI) bandpass filtering \cite{MicroDoppler}. Since respiration rate ($\sim$ 10-20 breaths per minute) and heart rate ($\sim$ 40-120 beats per minute) lie in non-overlapping intervals, LTI filtering appears to be a plausible approach. However, LTI filtering ignores the fact that the two components are not perfectly sinusoidal in shape, and that they thus have harmonics. When we apply a bandpass filter to keep components in the range [40,120] beats per minute, we (i) allow the harmonics of respiration to be part of the heart activity estimate, (ii) lose the higher harmonics of heart activity. Due to (ii), we end up with a smoother signal, and lose the `impulse-train' like appearance in the time domain, making the exact peak locations more ambiguous (see Fig.~\ref{fig:result}b). Due to (i), the shape of the heart actitivy estimate may be slightly altered, and this introduces further error when locating the peaks. Because of these reasons, we assert that LTI filtering is not ideal for this problem.

An interesting thread of research with a similar target can be collected under `morphological component analysis', or `resonance based signal processing' - see e.g.,  \cite{fad10p983,sta04p} for an overview of `morphological component analysis' for images, and \cite{nin14p156, sel14p596, bay14p95} for applications to 1D signals. These frameworks assume that components can be parsimoniously represented in distinct bases/frames. Concatenating the frames for the different components and looking for the sparsest solution (possibly with structure) leads to the sought decomposition. In our problem, however, we do not have a viable model for either component, other than the vague statement of smoothness for the high magnitude component.  In our experiments, resonance based processing did not yield good results, possibly because we were not able to find significantly distinct frames that can parsimoniously represent the individual components.

Another potentially useful framework for this problem is Empirical mode decomposition (EMD) \cite{hua98p903}. EMD aims to decompose a signal into `intrinsic mode functions' (IMFs) such that each IMF has a unique time-varying frequency in the time-frequency plane. EMD is originally defined through an algorithm \cite{hua98p903}, but other interpretations with alternative formulations/algorithms for extracting the components have been proposed \cite{hua13p174, pus14p313}. EMD has the potential to alleviate the leakage of the respiration harmonics into the heart actitvity estimate, that is mentioned to be an issue for LTI filtering. However, in practice, out-of-the-shelf EMD does not say which IMF belongs to respiration or heart activity, and is thus not straightforward to use. We also found that the components produced by EMD may be contaminated by a mixture of respiration and heart activity (see Fig.~\ref{fig:IMFs}).

\renewcommand{\sc}{0.35}
\begin{figure}
\centering
\includegraphics[scale = \sc]{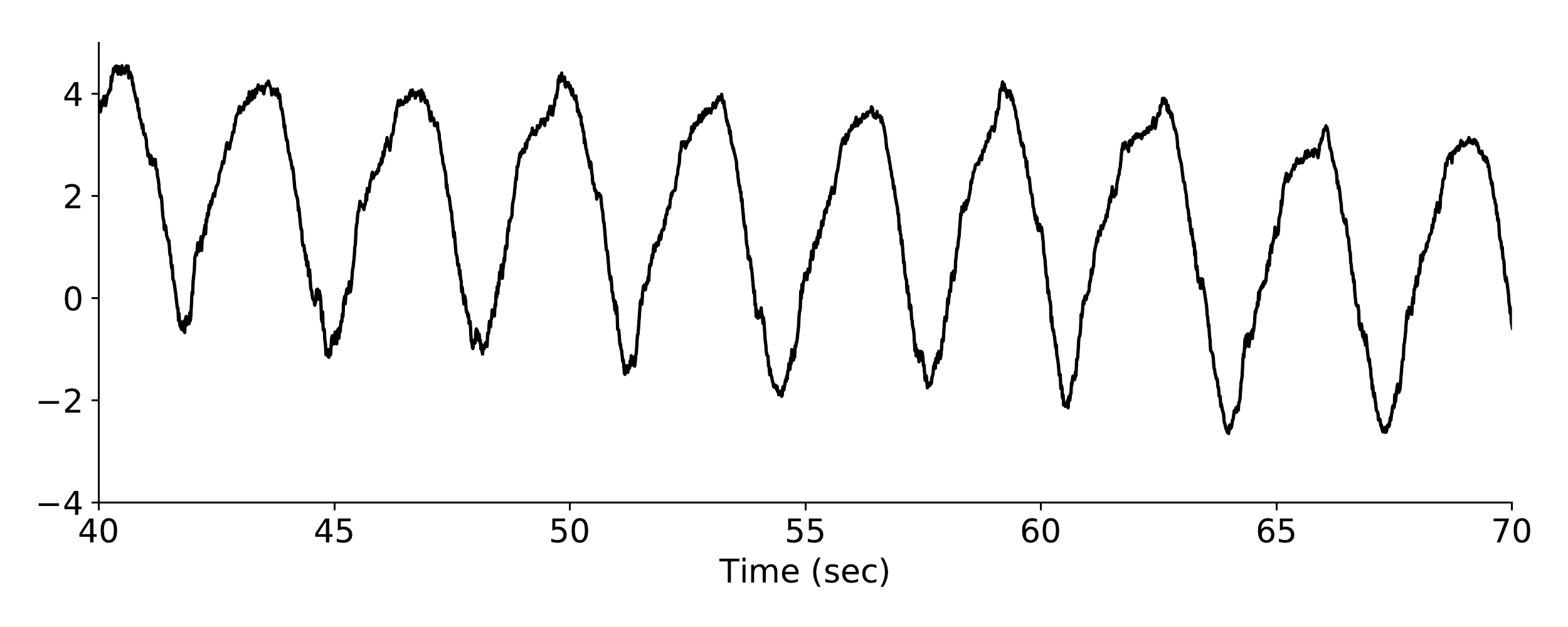}
\caption{Observed `vital signs' signal using radar, consisting of the sum of respiration and heart activity components.\label{fig:radar} }
\end{figure}
\subsection*{Proposed Method}
\begin{figure}
\centering
\scriptsize
\textsf{(a) Smooth and Transient Components}\\
\includegraphics[scale = \sc]{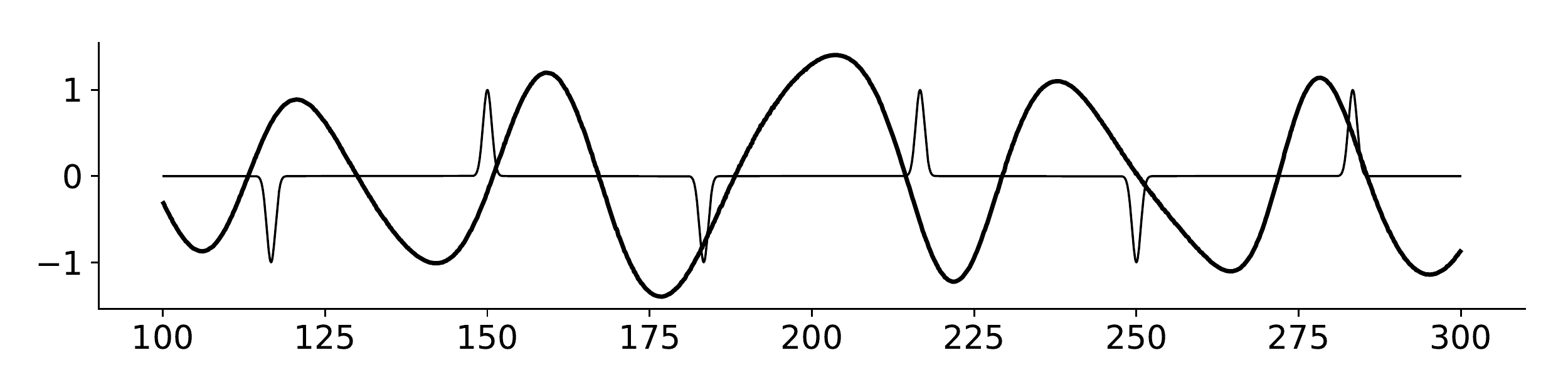}\\
\textsf{(b) Observed Composite Signal}\\
\includegraphics[scale = \sc]{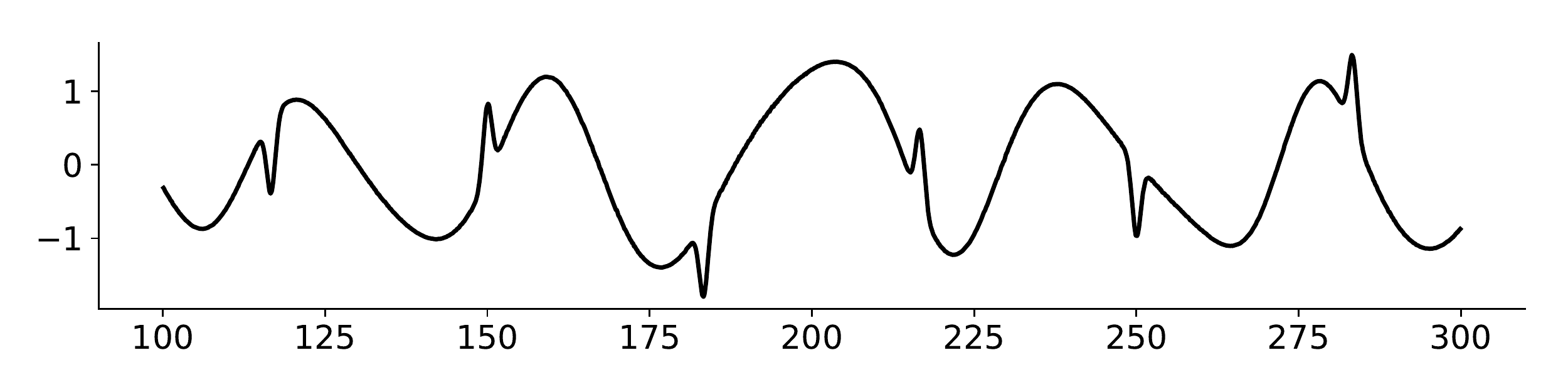}
\caption{A simple example to demonstrate the idea. (a) The smooth (thick line) and transient (thin line) component signals, making up the observed composite signal in (b).\label{fig:ideal}}
\end{figure}

\begin{figure}
\centering
\includegraphics[scale = \sc]{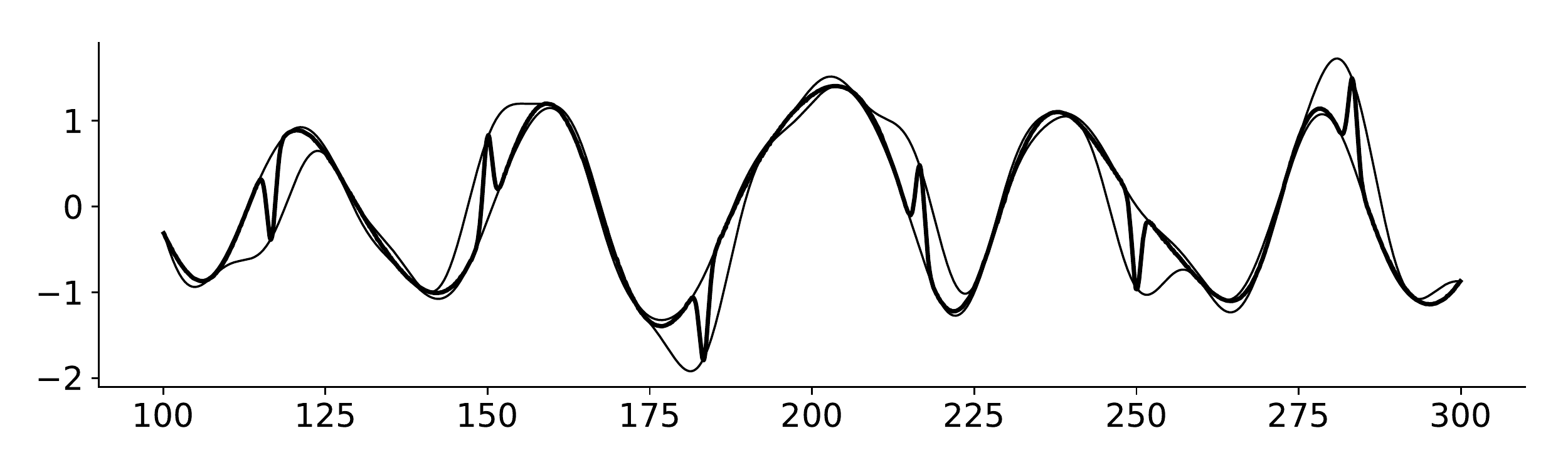}
\caption{Given the composite observation (thick line), we fit upper and lower envelopes (thin lines), that are as smooth as possible. The smoothness of the envelopes is disrupted by the perturbations in the observation.\label{fig:env} }
\end{figure}

\begin{figure}
\centering
\scriptsize
\textsf{(a) Estimation of the Smooth Component}\\
\includegraphics[scale = \sc]{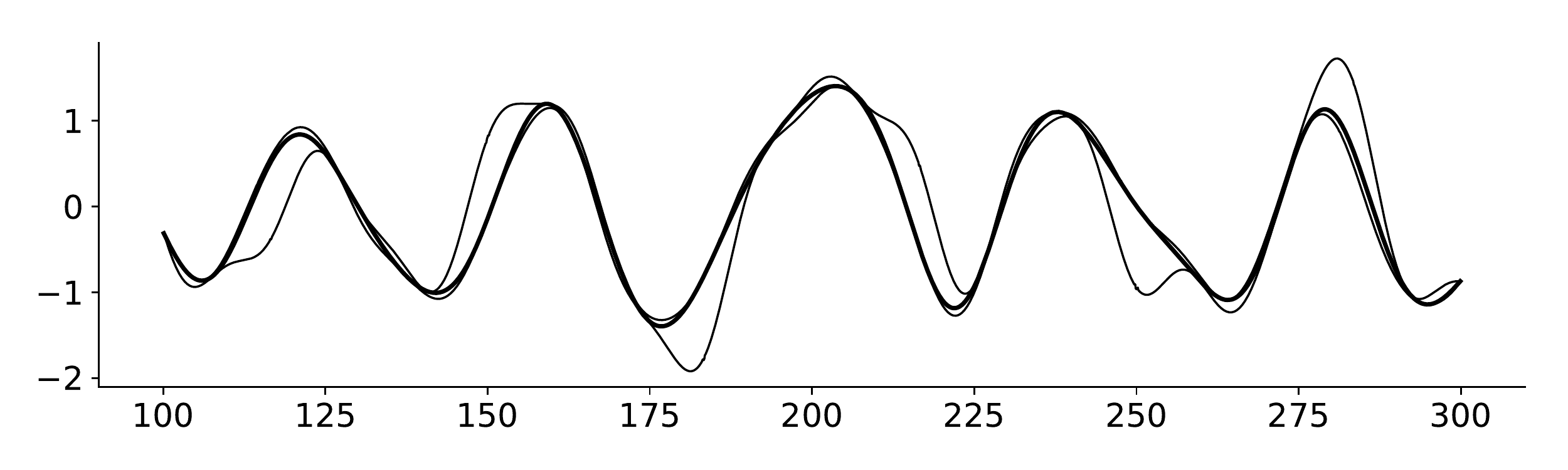}\\
\textsf{(b) Transient Component -- Estimated vs True}\\
\includegraphics[scale = \sc]{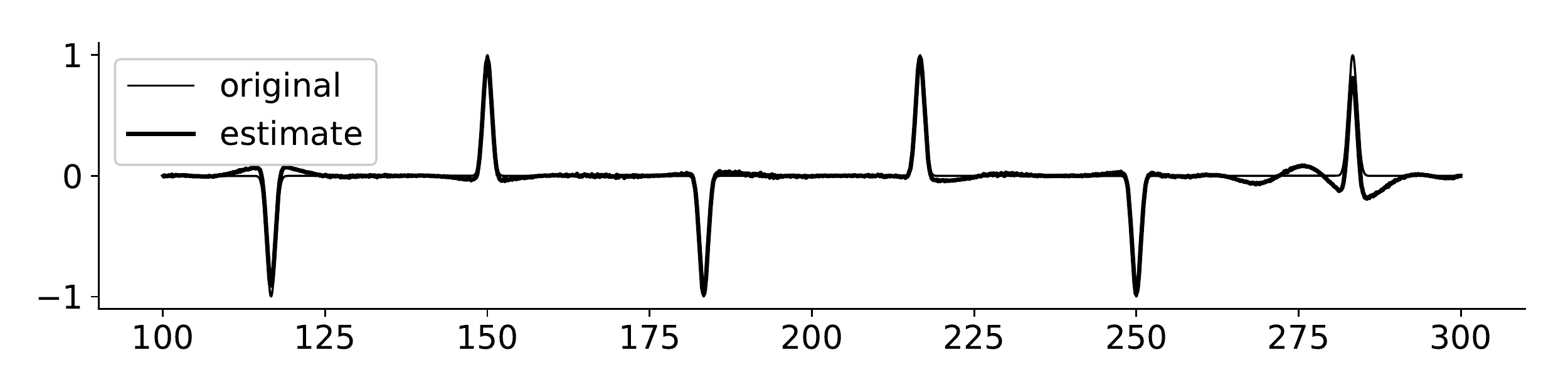}
\caption{(a) Given the envelopes,  the `smoothest' signal (thick line) that lies between the envelopes  (thin lines) forms our estimate of the smooth component. (b) Subtracting the smooth component estimate from the composite observation gives the estimate of the transient. \label{fig:estimate}}
\end{figure}

We pose the problem as decomposing a signal into smooth and transient components.
Let us outline our proposed approach on a toy example. We get to observe the composite signal in Fig.~\ref{fig:ideal}b, and would like to recover the components in Fig.~\ref{fig:ideal}a. As in the first step of EMD, we fit upper and lower envelopes to the observed signal. However, unlike in EMD, these envelopes are chosen to snugly sandwich the signal. The upper (lower) envelope is selected such that it is (i) as smooth as possible,  (ii) nowhere less (greater) than the observed signal, and (iii) as close as possible to the original signal (see  Fig.\ref{fig:env}). Note that the gap between the envelopes is reduced if the transient signal magnitude is small. Given the envelopes, the smooth component is estimated as the `smoothest signal' that lies between the envelopes (see Fig.~\ref{fig:estimate}). In seeking the smoothest signal, we no longer require the estimate to be close to the original composite signal strictly, as the envelopes already contain information about the composite signal.

The proposed method may be interpreted as LTI filtering under a nonlinear constraint (the output is constrained to lie between the envelopes). Since the envelopes are snug, their shape resembles that of the original smooth signal, except for disturbances caused by the transient signal. This in turn leads to a smooth signal estimate that preserves the harmonics (with the correct phase) of the underlying smooth signal. 

One feature of the proposed method we want to emphasize is that all of the steps are realized by solving an instance of a `constrained filtering problem' (to be detailed in Section~\ref{sec:formulation}). The constrained filtering problem to be introduced is a quadratic program, but amounts to applying a nonlinear operation on the input. By considering a specific dual of the problem, we make use of the problem's structure to devise a fast algorithm that is computationally favorable to direct off-the-shelf quadratic program solvers.

\subsection*{Outline}

We introduce and discuss the constrained filtering problem in Sec.~\ref{sec:formulation}. In Sec.~\ref{sec:reform}, we consider the dual of the minimization problem and derive an efficient algorithm for solving the dual problem.
Results of an experiment demonstrating the utility of the proposed algorithm on a real signal is described in Sec.~\ref{sec:exp}. Sec.~\ref{sec:conc} contains some remarks about noise.

\section{Formulation}\label{sec:formulation}
In this section, we introduce the `constrained filtering problem' mentioned in the Introduction, and 
discuss how to use it to achieve a decomposition.
\subsection{Constrained Filtering}\label{sec:basic}
Part of our problem requires us to estimate a  smooth signal. Gaussian processes are usually used as a prior distribution over smooth random signals, and are useful for deriving principled formulations \cite{GPML, cru13p40}. Specifically, suppose a signal of interest $x$ is modelled as a stationary  zero-mean Gaussian process with covariance
\begin{multline}
C_{\sigma}(x_n, x_m) = \epsilon\,\delta(n-m) \\ +  \begin{cases} 
 0, \text{ if } \exp\bigl( - (n - m)^2 / \sigma^2 \bigr) < \tau.\\
 \exp\bigl( - (n - m)^2 / \sigma^2 \bigr), \text{ otherwise}.
 \end{cases}
\end{multline}
where $\tau > 0$ is a threshold, $\epsilon$ is a small constant, and 
\begin{equation}
\delta(k) = \begin{cases} 1, &\text{if } k=0,\\
0, &\text{if } k\neq 0.
\end{cases}
\end{equation}
The addition of the term $\epsilon \delta(n-m)$ ensures that the covariance matrix $C_{\sigma}$ is positive semi definite if $\epsilon$ is sufficiently large -- see \cite{storkey99ICANN} for a further discussion, and alternatives to ensure positive definiteness. In practice, we found that the value of $\epsilon$ needed to make $C_{\sigma}$ is very small and the dual of the problem, which we discuss later has a form that allows us to ignore $\epsilon$ altogether, without causing any numerical instability, and significant bias.
Note now that, if $|n-m|$ is large, then  $C_{\sigma}(x_n, x_m)=0$.
Therefore, for $x \in \R^N$, $C_{\sigma}$ is an $N\times N$ Toeplitz matrix. Using  $C_{\sigma}$, let us define $S_{\sigma}(x) = x^T\, C_{\sigma}^{-1}\, x$.

If $y$ denotes noisy observations of a smooth signal, a denoising formulation based on Gaussian processes could be,
\begin{equation}\label{eqn:Gaussian}
\arg \min_x\,\frac{1}{2}\| y - x \|_2^2 + S_{\sigma}(x).
\end{equation}
This formulation coincides with that of maximum a posteriori (MAP) estimation \cite{Kay1}, where the first term is the likelihood, provided the noise is standard Gaussian. The solution of \eqref{eqn:Gaussian} is $(I + \,C_{\sigma}^{-1})^{-1}\,y$. If $y$ were an infinite length discrete-time signal, this operation would be equivalent to LTI lowpass filtering with a kernel determined by $C_{\sigma}$. For finite-length $y$, this operation is no longer LTI filtering exactly, but only approximately.

Consider now the following variation on \eqref{eqn:Gaussian}:
\begin{equation}\label{eqn:basicform}
\min_x\,\frac{\lambda}{2}\| y - x \|_2^2 + S_{\sigma}(x) \text{ subject to } a_i \leq x_i \leq b_i,
\end{equation}
where $y \in \R^N$ is an observation, $\lambda \in \R_+$ is a weight parameter, $a_i < b_i$'s are given constants. We denote the minimizer of \eqref{eqn:basicform} as $\hat{x}_{y,\lambda, \sigma, a, b}$. This problem seeks the `smoothest' signal in a given interval, that is close to $y$. Relying on our previous interpretation of \eqref{eqn:Gaussian}, we regard the mapping $y \to \hat{x}_{y,\lambda, \sigma, a, b}$ as a constrained filtering operation. 

The problem \eqref{eqn:basicform} is simple but flexible enough to realize all of the steps of the proposed method, outlined in the Introduction. Specifically, by setting $a = y$, $b = \infty$, we can obtain an upper envelope, $u$. By setting $a = -\infty$, $b = y$, we obtain a lower envelope, $\ell$. Finally, setting $a = \ell$, $b = u$, we obtain the estimate of the smooth component. Subtracting the smooth component from the composite observation, we obtain the transient component. 

We next discuss  briefly how to set the parameters in \eqref{eqn:basicform}.

\subsection{Parameters of the Formulation}
We expect different behavior from the envelopes and the smooth signal estimate. 
 In order to make the envelopes fit tightly, we can increase the value of $\lambda$, or penalize deviation from smoothness less by reducing $\sigma$. On the other hand, once we have the envelopes, to reduce the influence of the underlying observation $y$, we reduce $\lambda$, and possibly increase $\sigma$, to obtain a smoother signal.

These considerations lead to Algorithm~\ref{algo:smooth}.

\begin{algorithm}\caption{Smooth Component Estimation}\label{algo:smooth}
\begin{algorithmic}[1]
\Require Input signal $y$
\State Set $0<\lambda_1 \ll \lambda_0$, $0 < \sigma_0 \leq \sigma_1$
\State $\ell \gets \hat{x}_{p}$ for $p = \{y, \lambda_0, \sigma_0, \min(y), y\}$  \texttt{\footnotesize \%lower env.}
\State $u \gets \hat{x}_p$ for $p = \{y, \lambda_0, \sigma_0, y, \max(y)\}$  \texttt{\footnotesize \%upper env.}
\State $x^* \gets \hat{x}_p$ for $p = \{y, \lambda_1, \sigma_1, \ell, u\}$  \texttt{\footnotesize \%smooth component}
\State $t^* \gets y - x^*$    \texttt{\footnotesize \%transient component}
\end{algorithmic}
\end{algorithm}

Algorithm~\ref{algo:smooth} assumes we know how to solve \eqref{eqn:basicform} to obtain $\hat{x}_p$ for a given set of parameters $p$. 
We next discuss how to efficiently solve \eqref{eqn:basicform}.

\section{Reformulating the Problem}\label{sec:reform}
The problem \eqref{eqn:basicform} is a quadratic program \cite{Boyd}. Direct approaches to this problem require multiplications with $C_{\sigma}^{-1}$ (see e.g., Chp. 16 of \cite{Nocedal}).  Unfortunately, the Toeplitz structure of $C_{\sigma}$ is lost during inversion, and $C_{\sigma}^{-1}$ is not available in closed form. Further, even if we had  $C_{\sigma}^{-1}$, lack of structure prevents us to realize multiplication with  $C_{\sigma}^{-1}$ efficiently. 

To exploit the Toeplitz structure of  $C_{\sigma}$, we consider a dual of \eqref{eqn:basicform}, and derive a splitting algorithm that uses FFTs.

\subsection{A Dual Problem}
Let us denote the constraint set as $B = \{ x : a_n \leq x_n \leq b_n\}$. We can write \eqref{eqn:basicform} as
\begin{equation}\label{eqn:primal}
\min_x \, \frac{\lambda}{2}\| y - x \|_2^2 + \frac{1}{2}\,x^T\,C_{\sigma}^{-1}\,x + i_B(x),
\end{equation}
where $i_B(\cdot)$ is the indicator function of $B$, defined as $i_B(x) = 0$ if $x\in B$, and $i_B(x) = \infty$ if $x\notin B$ \cite{HiriartFund}.
Using the Fenchel dual of the quadratic \cite{HiriartFund, Boyd},  we express \eqref{eqn:primal} as
\begin{equation}\label{eqn:saddle}
\min_x \, \max_z \frac{\lambda}{2}\| y - x \|_2^2  + i_B(x) + \langle z,x \rangle - \frac{1}{2}\,z^T\,C_{\sigma}\,z,
\end{equation}
Changing the order of $\min / \max$ and solving for $x$, we find
\begin{equation}\label{eqn:proj}
x = P_B(y - z/\lambda),
\end{equation}
where $P_B(\cdot)$ is the projection operator onto $B$. 
We plug \eqref{eqn:proj} in \eqref{eqn:saddle}, and 
rearrange terms to obtain a dual problem as
\begin{multline}\label{eqn:dual}
\min_z\, \frac{\lambda}{2}\langle 2(y-z/\lambda) - P_B(y-z/\lambda), P_B(y-z/\lambda) \rangle \\+ \frac{1}{2}\,z^T\,C_{\sigma}\,z.
\end{multline}
If $z$ solves \eqref{eqn:dual}, then \eqref{eqn:proj} solves the primal problem \eqref{eqn:primal}.

We also note that solving \eqref{eqn:saddle} for $z$, we find 
\begin{equation}\label{eqn:cond2}
x = C_{\sigma}\,z.
\end{equation}
Optimality conditions for checking convergence are provided in Appendix~\ref{sec:Optimality}.

\subsection{A Modified Problem}
We will obtain a fast algorithm for \eqref{eqn:dual} by adapting the Douglas-Rachford (DR) algorithm \cite{combettes_chp, Bauschke, eck92p293}. In principle, any convex splitting algorithm can be used for tackling \eqref{eqn:primal} or \eqref{eqn:dual}  \cite{combettes_chp, Bauschke, cha11p120}. We opt for the Douglas-Rachford algorithm because its steps involve solutions of non-trivial problems, but can be efficiently realized in our case, and it does not require variable splitting. 

A crucial ingredient for DR is the proximity operator \cite{Bauschke, combettes_chp}.
\begin{defn}
For a convex $g:\R^n \to \R$, and $\alpha > 0$, the proximity operator for $g$, namely $J_{\alpha\,g}(\cdot) : \R^n \to \R^n$, is defined as
\begin{equation}
J_{\alpha\,g}(z) = \arg \min_x\, \frac{1}{2} \| x - z \|_2^2 + \alpha\,g(x). 
\end{equation}
\end{defn}
For a convex problem involving the sum of two functions,
\begin{equation}\label{eqn:split}
\min_x\, f(x) + g(x),
\end{equation}
the DR iterations are of the form \cite{Bauschke}
\begin{equation}\label{eqn:DR}
u^{n} = \gamma\, u^{n-1} + (1-\gamma)\, \Bigl(N_{\alpha g} \bigl(N_{\alpha f}(u^{n-1}) \bigr) \Bigr),
\end{equation}
for $0 < \gamma < 1$, where $N_{\alpha\,f} (\cdot) := (2\,J_{\alpha f} - I)(\cdot)$, $N_{\alpha\,g} (\cdot) := (2\,J_{\alpha g} - I)(\cdot)$.
This sequence converges to a point $u$ such that $x = J_{\alpha f}\,(u)$ solves \eqref{eqn:split}.

In order to employ DR, we need to split the function in \eqref{eqn:dual}. However, straightforward splitting requires applying $(I + \alpha C_{\sigma})^{-1}$ at each iteration, and we face the same issue of inverting $C_{\sigma}$. In order to avoid this, we consider a modified problem, following the main idea in \cite{bay15p264}.

Specifically, we note that $C_{\sigma}$ can be embedded in a larger \emph{circulant} matrix $\tilde{C}_{\sigma}$ (see Sec.V in \cite{bay15p264} for an  example) as
\begin{equation}\label{eqn:embed}
\tilde{C}_{\sigma} = \begin{bmatrix}
C_{\sigma} & D_0\\
D_1 & D_2
\end{bmatrix},
\end{equation}
for some $D_i$'s,  where the sizes of smallest $D_i$'s are determined by the band-size of the Toeplitz $C_{\sigma}$. 
Notice now that
\begin{equation}
z^T\, C_{\sigma}\, z = \begin{bmatrix}z \\ \tilde{z}  \end{bmatrix}^T\,\tilde{C}_{\sigma}\, \begin{bmatrix} z \\ \tilde{z}  \end{bmatrix}, \text{ for }\tilde{z} = 0.
\end{equation}
This motivates the following modification of \eqref{eqn:dual},
\begin{multline}\label{eqn:modified}
\min_{z, \tilde{z}}\, \Bigl\{ h(z, \tilde{z}) :=  \frac{1}{2}\,\begin{bmatrix} z \\ \tilde{z} \end{bmatrix}^T\,\tilde{C}_{\sigma}\,\begin{bmatrix} z \\ \tilde{z} \end{bmatrix} +  i_0(\tilde{z}) \\ +\frac{\lambda}{2}\langle 2(y-z/\lambda) - P_B(y-z/\lambda), P_B(y-z/\lambda) \rangle \Bigr\},
\end{multline}
where $i_0(\tilde{z})$ is the indicator function for the set  $\{0\}$. 

The following proposition is a consequence of the development so far.
\begin{prop}
If $(z^*,\tilde{z}^*)$ is a solution to \eqref{eqn:modified}, then $x^* = P_B(y - z^*/\lambda) = C_{\sigma}\,z^*$ is a solution to \eqref{eqn:basicform}.
\end{prop}

The DR algorithm on a specific splitting of this cost function leads to Algorithm~\ref{algo:pseudo}. The derivation is provided in Appendix~\ref{sec:DR}. 
Even though this algorithm addresses a problem with more variables than  \eqref{eqn:basicform}, the capability to exploit the Toeplitz structure makes up for the increase in dimension -- see \cite{bay15p264} for a discussion.
\begin{algorithm}\caption{Computation of $\hat{x}_{y,\lambda, \sigma, a, b}$ that solves \eqref{eqn:basicform}}\label{algo:pseudo}
\begin{algorithmic}[1]
\State Set $0 < \gamma < 1$, $0 < \alpha$. 
\State Given $N \times N$ Toeplitz $C_{\sigma}$, find the smallest $(N + K) \times (N+K)$ circulant  $\tilde{C}_{\sigma}$ s.t. \eqref{eqn:embed} holds
\State Initialize $u\in \R^N$, $\tilde{u} \in \R^K$.
\State $c \gets \lambda \bigl( y - (1 + \alpha / \lambda) a \bigr)$
\State $d \gets \lambda \bigl( y - (1 + \alpha / \lambda) b \bigr)$
\Repeat
\State $\begin{bmatrix} t \\ \tilde{t} \end{bmatrix} \gets \Bigl( 2 (I + \alpha\,\tilde{C}_{\sigma})^{-1} - I \Bigr)\,\begin{bmatrix} u \\ \tilde{u} \end{bmatrix}$ \texttt{\footnotesize \%using FFT}
\State $t_n \gets \begin{cases}
t_n + 2\,\alpha\,b_n,  \text{ if } t_n <d_n,\\
\dfrac{2 \alpha\, y_n + (1- \alpha / \lambda)\,t_n}{1 + \alpha / \lambda},   \text{ if } c_n \leq t_n  \leq d_n,\\
t_n + 2\, \alpha\,a_n,   \text{ if }  c_n < t_n,
\end{cases}$ for  $n=1,\ldots,N$
\State $u \gets \gamma \, u + (1 - \gamma)\, t$
\State $\tilde{u} \gets \gamma \, \tilde{u} - (1 - \gamma)\, \tilde{t}$
\Until{some convergence criterion is met}
\State $\begin{bmatrix} z \\ \tilde{z} \end{bmatrix} \gets (I + \alpha\,\tilde{C}_{\sigma})^{-1} \,\begin{bmatrix} u \\ \tilde{u} \end{bmatrix}$ \texttt{\footnotesize \%using FFT}
\State $\hat{x}_{y,\lambda, \sigma, a, b} \gets P_B(y - z / \lambda)$
\end{algorithmic}
\end{algorithm}

\section{Demonstration  of the Method on Real Data}\label{sec:exp}

We now consider a vital signs monitoring experiment with real data, obtained from radar.
The subject was stationary during signal acquisition, and kept a steady breathing pattern for about five minutes. An excerpt from the phase signal from radar is shown in Fig.~\ref{fig:radar}. We compare the result of applying the proposed algorithm, bandpass filtering, as well as EMD to this signal. Some of the details of the experiment can be found in Appendix~\ref{sec:exp_det}.

\begin{figure}
\centering
\footnotesize
\includegraphics[scale = \sc]{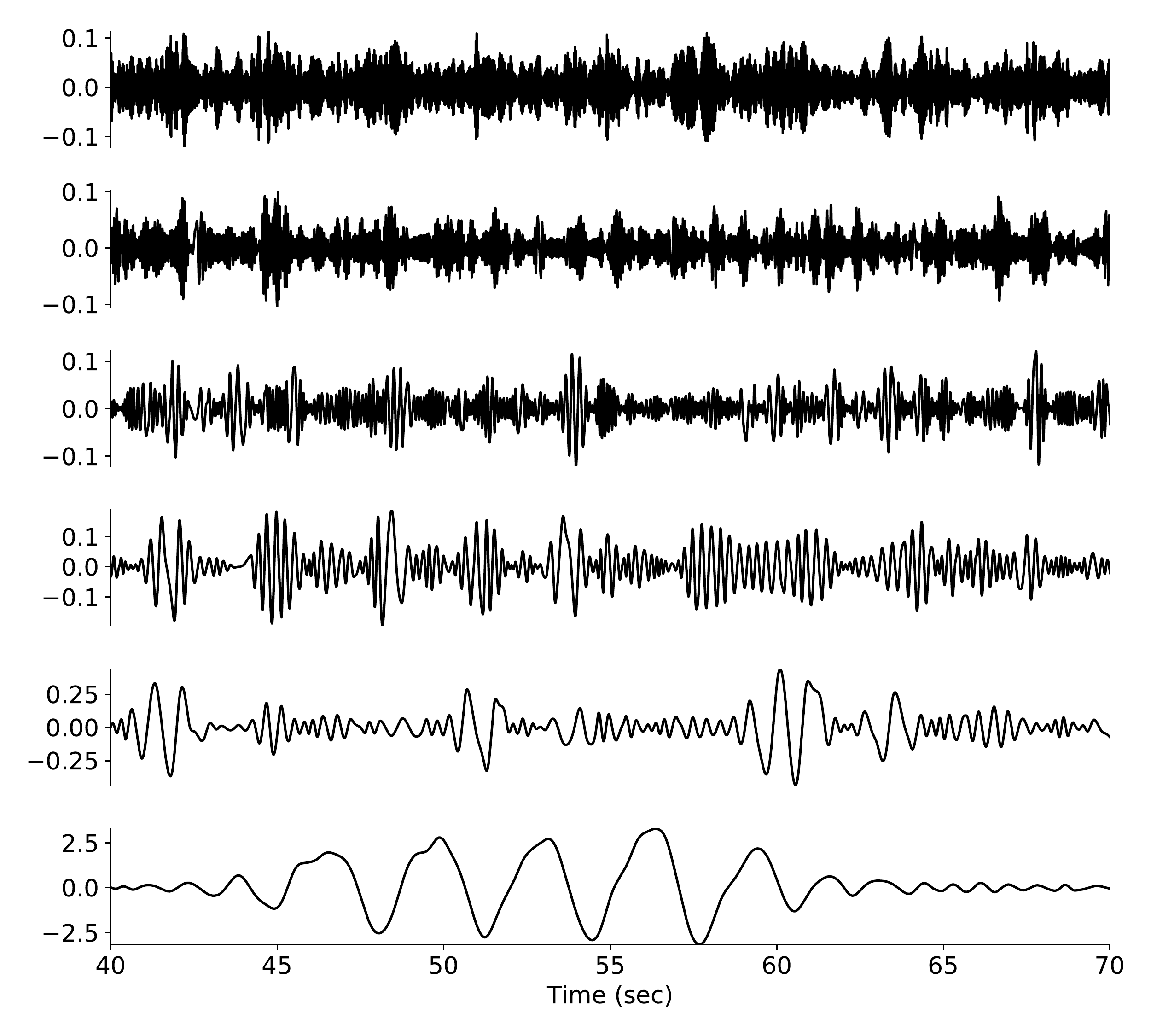}
\caption{The first 6 of the 11 intrinsic mode functions obtained by applying EMD to the observed signal from radar. \label{fig:IMFs}}
\end{figure}

\begin{figure}
\centering
\scriptsize
\textsf{(a) Empirical Mode Decomposition}
\includegraphics[scale = \sc]{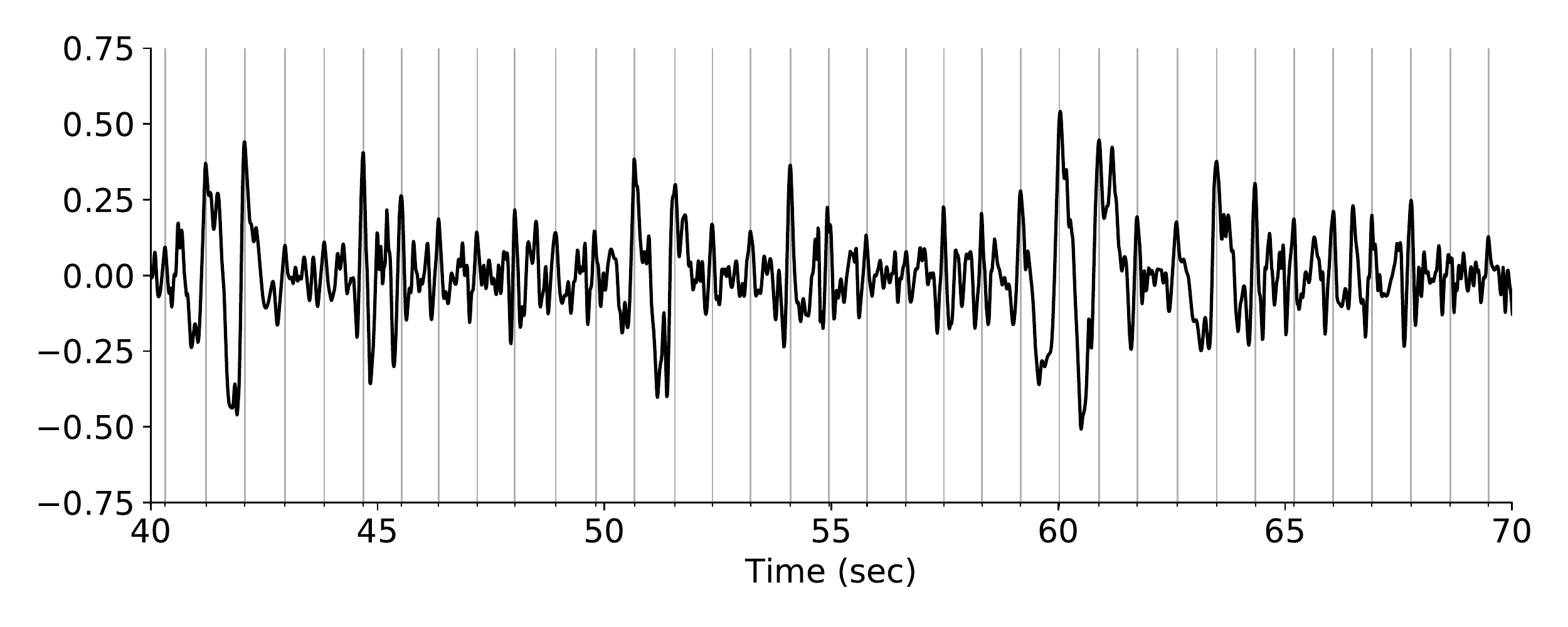}\\
\textsf{(b) Bandpass filtering}\\
\includegraphics[scale = \sc]{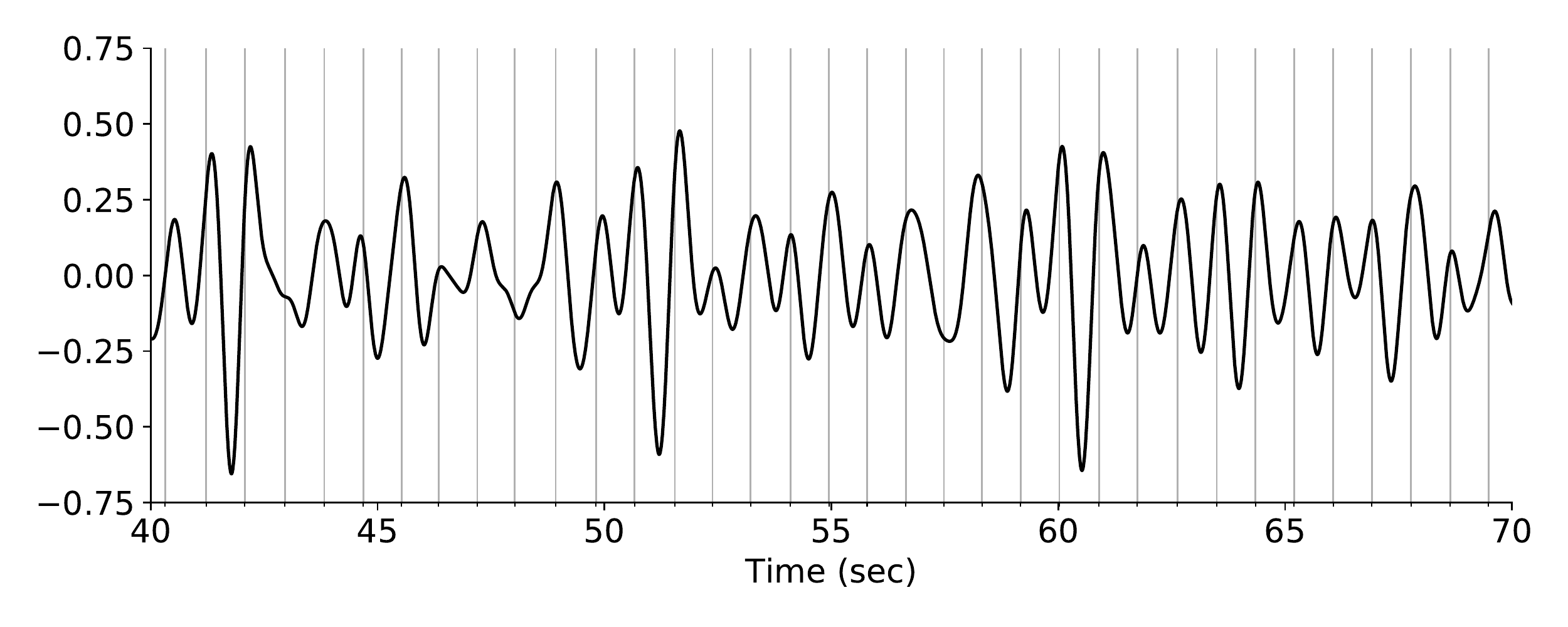}\\
\textsf{(c) Proposed Method}\\
\includegraphics[scale = \sc]{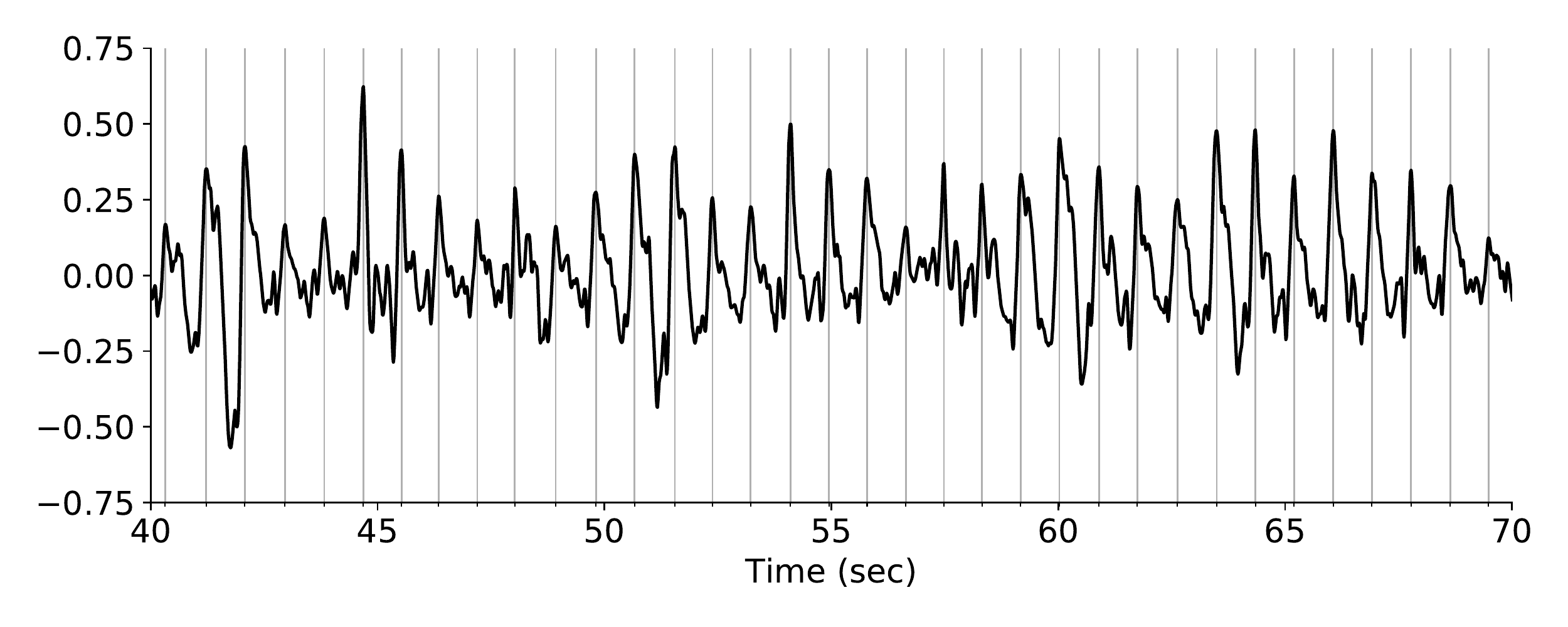}

\caption{Estimates of heart activity obtained using (a) EMD, (b) bandpass filtering, and (c) the proposed method. Vertical bars indicate the local peaks of the signal in (c).\label{fig:result}}
\end{figure}

The first six IMFs from EMD are shown in Fig.~\ref{fig:IMFs}\footnote{We used EMD code by G. Rilling, associated with \cite{ril03EMD}.}. The remaining IMFs all have higher amplitudes, and slow variation, and therefore are not correlated with heart activity. Note that, no single IMF really captures a plausible heart activity. In order to estimate a heart activity signal, we consider the sum from second up to fifth IMF, because the first IMF is noise like, and the sixth IMF contains a high amplitude segment which cannot be coming from heart activity. The resulting estimate of heart activity from EMD is shown in Fig.~\ref{fig:result}a. 

Fig~\ref{fig:result}b shows the output of a linear phase bandpass filter applied to the input. Finally, we use the proposed method to obtain a smooth signal, and subtract it from the composite signal, to obtain the estimate of the heart activity signal shown in Fig~\ref{fig:result}c. 

In Fig.~\ref{fig:result}a,b,c, vertical bars indicate the peaks of the heart activity signal obtained with the proposed method. We see that, occasionally, the EMD estimate is in sync with the proposed method, but in general, the behavior of the EMD estimate is not consistent. Bandpass filtering produces a fairly reasonable estimate, but it occasionally misses beats, and its peaks deviate around those of the proposed method. Overall, we found that the standard deviation of beat to beat intervals obtained from the bandpass filtered estimate is much larger than is typical. Therefore, we are led to believe that the deviation of the peaks mentioned above is due to the additional bias coming from the harmonics of respiration.

\section{Discussion}\label{sec:conc}
One issue we have not addressed is noise in the observed composite signal. Lacking a further model on noise, the proposed method is likely to include noise in the estimate of the transient component. Therefore, we recommend denoising as a preprocessing stage. As long as the transient component is not buried in noise, we expect the method to provide valuable estimates, as demonstrated in the experiment with a real signal.

\section*{Acknowledgement}
We thank Sundar Palani, Analog Devices Inc., for his help in obtaining the radar signal used in this paper. We also thank the anonymous reviewers for their comments and suggestions.

\appendix

\subsection{A Set of Optimality Conditions for Checking Convergence}\label{sec:Optimality}
A pair $(x,z)$ is a saddle point of  \eqref{eqn:saddle} if and only if \eqref{eqn:proj} and \eqref{eqn:cond2} hold.
Combining these two equations, we find that $z$ solves the dual problem \eqref{eqn:dual} if and only if 
\begin{equation}\label{eqn:fixedz}
C_{\sigma}\,z = P_B(y-z /\lambda).
\end{equation}
Similarly, $x$ solves the primal problem \eqref{eqn:primal} if and only if
\begin{equation}\label{eqn:fixedx}
x = P_B(y- C_{\sigma}^{-1}\,x /\lambda).
\end{equation}

Fixed point iterations can be derived from \eqref{eqn:fixedx} and \eqref{eqn:fixedz} to solve the primal and dual problem respectively. 

\subsection{Derivation of Algorithm~\ref{algo:pseudo}}\label{sec:DR}
For DR iterations, we split $h(z,\tilde{z})$ in \eqref{eqn:modified} as follows
\begin{multline}\label{eqn:fg}
f(z,\tilde{z}) = \frac{1}{2}\,\begin{bmatrix} z \\ \tilde{z} \end{bmatrix}^T\,\tilde{C}_{\sigma}\,\begin{bmatrix} z \\ \tilde{z} \end{bmatrix},\\
g(z,\tilde{z}) =  i_0(\tilde{z})  \\  + \frac{\lambda}{2}\langle 2(y-z/\lambda) - P_B(y-z/\lambda), P_B(y-z/\lambda) \rangle .
\end{multline}
We need to find expressions for the operators  $N_{\alpha\,f} (\cdot)$,  $N_{\alpha\,g} (\cdot)$.
First, note that,
\begin{align}
\begin{bmatrix} t \\ \tilde{t} \end{bmatrix} &:= N_{\alpha\,f} (u, \tilde{u}) = 2J_{\alpha f} (u,\tilde{u})  - \begin{bmatrix} u \\ \tilde{u} \end{bmatrix}\\
&=  \Bigl( 2\,(I+ \alpha\,\tilde{C}_{\sigma})^{-1} - I \Bigr)\, \begin{bmatrix} u \\ \tilde{u} \end{bmatrix}. 
\end{align} 
Thanks to the circulant structure of $\tilde{C}_{\sigma}$, this can be evaluated efficiently using FFTs \cite{bay15p264}.

Let us now consider the second operator $N_{\alpha\,g} (\cdot) =  (2\,J_{\alpha g} - I)(\cdot)$. Even though $g(\cdot)$ has a relatively complicated expression, all of the variables are decoupled. First, for $\begin{bmatrix} v \\ \tilde{v} \end{bmatrix} = N_{\alpha\,g} (t, \tilde{t})$, 
we have $\tilde{v} = - \tilde{t}$.

To find an expression for $v$, we first derive an expression for the proximity operator $J_{\alpha g}(\cdot)$. Let 
\begin{multline}
\begin{bmatrix}
r\\
\tilde{r}
\end{bmatrix} := \Bigl\{ J_{\alpha g}\, (t, \tilde{t}) \\= \arg \min_{s, \tilde{s}}\, \frac{1}{2} \left\| \begin{bmatrix}
t\\
\tilde{t} \end{bmatrix} -  \begin{bmatrix}
s\\
\tilde{s} \end{bmatrix}\right\|_2^2 + \alpha \,g(s, \tilde{s}) \Bigr\}.
\end{multline}
We first note that $\tilde{r} = 0$. To obtain $r$, we need to solve
\begin{equation}
\min_s \, \sum_n \, \frac{1}{2}\, (t_n - s_n)^2 +\alpha\, \lambda\, q_n(y_n - s_n/\lambda),
\end{equation}
where, for $P_n(\cdot)$ denoting the projector onto the interval $[a_n, b_n]$, the function $q_n : \R \to \R$ is defined as
\begin{equation}
q_n(\cdot) = \frac{1}{2}\,(2\cdot - P_n(\cdot))\, P_n(\cdot)\\
\end{equation}
It follows that the minimization problem for $s$ is decoupled with respect to its entries, and $r_n$ depends only on $t_n$.

Let us now find the expression for $r_n$. First, note that
\begin{equation}
q_n(s) = \begin{cases}
\frac{1}{2}\,(2s - a_n)\,a_n, & \text{ if } s < a_n,\\
\frac{1}{2}\,s^2 & \text{ if } a_n \leq s \leq b_n,\\
\frac{1}{2}(2s - b_n)\,b_n & \text{ if } b_n < s.
\end{cases}
\end{equation}
The proximity operator of this function is
\begin{multline}\label{eqn:proxd}
J_{\alpha\,q_n}(s) = \\ \begin{cases}
s - \alpha\,a_n,  & \text{ if } s < (1 + \alpha)\,a_n,\\
s / (1 + \alpha),  & \text{ if } (1 + \alpha)\,a_n \leq s \leq (1 + \alpha)\,b_n,\\
s - \alpha\, b_n,  & \text{ if } (1 + \alpha)\,b_n < s.
\end{cases}
\end{multline}
In order to find the expression for $r_n$, we need the proximity operator for the function $\lambda\, q_n(y_n - \cdot / \lambda)$, which we obtain by a change of variables as
\begin{multline}
r_n = \arg \min_s\, \frac{1}{2}\,\| s - t_n\|_2^2 + \alpha\, \lambda \, q_n(y_n - s/\lambda)\\
=\lambda\Bigl( y_n -  \Bigl\{\arg \min_w\, \frac{\lambda^2}{2}\,\| w -  (y_n - t_n/\lambda) \|_2^2 + \alpha \lambda\, q_n(w) \Bigr\}\Bigr)\\
= \lambda\Bigl( y_n - J_{(\alpha\,/\lambda)\, q_n}(y_n - t_n/\lambda) \Bigr)\label{eqn:exprJd}.
\end{multline}
Plugging \eqref{eqn:proxd} in \eqref{eqn:exprJd}, the expression for $r_n$ is
\begin{equation}\label{eqn:reflected2}
 \begin{cases}
t_n + \alpha\,b_n,  \text{ if } t_n < \Bigl\{ d_n := \lambda \Bigl( y - (1 + \alpha / \lambda) b_n \Bigr) \Bigr\},\\
t_n + \alpha\,a_n,   \text{ if } t_n > \Bigl\{ c_n := \lambda \Bigl( y - (1 + \alpha / \lambda) a_n \Bigr) \Bigr\},\\
\dfrac{1}{1 + \alpha / \lambda} (\alpha y_n + t_n),   \text{ if } c_n \leq t_n  \leq d_n.
\end{cases}
\end{equation}
Finally, for $\begin{bmatrix} v \\ \tilde{v} \end{bmatrix} = N_{\alpha\,g} (t, \tilde{t})$, the expression for  $v_n = 2 r_n - t_n$ is given by (for $c$, $d$ defined as in \eqref{eqn:reflected2})
\begin{equation}
\begin{cases}
t_n + 2\,\alpha\,b_n,  \text{ if } t_n < d_n,\\
t_n + 2\, \alpha\,a_n,   \text{ if } t_n > c_n,\\
\dfrac{1}{1 + \alpha / \lambda} \bigl(2 \alpha y_n + (1- \alpha / \lambda)\,t_n\bigr),   \text{ if } c_n \leq t_n  \leq d_n.
\end{cases}
\end{equation}

\subsection{Notes on the Experiment in Section~\ref{sec:exp}}\label{sec:exp_det}
The sampling frequency for the signal was $10^3 / 12$. For the bandpass filter, we used a truncated sinc filter with 5000 taps, with a passband of $[40,120]$ beats per minute.

For the proposed method, one trick we found useful is to add and subtract a bias signal $b$, when computing the envelopes. That is, in step-2 of Algorithm~\ref{algo:smooth}, we compute $\ell =b +  \hat{x}_p$ for $p = \{y - b, \lambda_0, \sigma_0, \min(y), y - b \}$, where $y - b \leq 0$. This ensures a tighter envelope. Similarly, for $u$ we use a $b$ such that $y - b \geq 0$, and set $u = b +  \hat{x}_p$ for $p = \{y - b, \lambda_0, \sigma_0, y-b, \max(y)\}$.

Another useful trick, especially for signals with trend is to fit coarse lower and upper envelopes (using the constrained filtering problem with very high values of $\sigma$). The trend is then estimated as the average of the envelopes. These coarse upper and lower envelopes can also be used to normalize the magnitude of the signal.

\subsection{A Synthetic Experiment}\label{sec:sexp}

In this section, we demonstrate the algorithm on a random selection of smooth/transient signals.  An instance of the smooth component, transient component, and the observation (smooth + transient) is shown in Fig.~\ref{fig:Exp}a, b, c. Notice that the magnitude of the transient component is much smaller than the smooth component.

The smooth component is produced by randomly time-warping a sinusoid, and multiplying the warped sinusoid with a smoothly varying magnitude function. The transient component is obtained by passing a sample from a Gaussian process through a nonlinearity. All of the random functions involved are initialized by sampling a Gaussian process. Specifically, we work with a zero-mean, stationary GP with covariance
\begin{equation}
\mathbb{E}\Bigl( x(t), x(t') \Bigr) = c_0\, \exp\Bigl( - (t-t')^2 / c_1\Bigr) + c_2 \, I.
\end{equation}
The warped time-variable is obtained as
\begin{equation}
\tilde{t} = t + s(t),
\end{equation}
where $s(t)$ is a sample from the GP with parameters $c_0 = 25$, $c_1 = 500$, $c_2 = 10^{-3}$. $s(t)$ is sampled on a uniform grid where the sampling frequency for the grid is $f_s = 10$ Hz. Given the warped time variable, the smooth component is produced as
\begin{equation}
x(t) = \cos\bigl(0.5 \pi\,\tilde{t}\bigr) \cdot \bigl(0.05 \, m(t) + 1 \bigr),
\end{equation}
where $m(t)$ is a sample from the GP with parameters $c_0 = 25$, $c_1 = 2500$, $c_2 = 5\cdot10^{-4}$. 
The transient signal $y(t)$ is defined by passing a GP through a pointwise nonlinearity as
\begin{equation}
y(t) = Q\Bigl( f(t)\Bigr),
\end{equation}
where $f(t)$ is a sample from the GP with parameters $c_0 = 0.1$, $c_1 = 10$, $c_2 = 10^{-5}$, and $Q(\cdot)$ denotes the nonlinear function defined as,
\begin{equation}
Q(u) = \begin{cases}
u, & \text{if } |u| > 1,\\
u^2, & \text{if } 0\leq u \leq 1,\\
-u^2, & \text{if } -1\leq u \leq 0.
\end{cases}
\end{equation}

We note that even though the components are Gaussian processes, the smooth/transient components are no longer Gaussian processes.

Given an observation $z = x + y$, the proposed method aims to primarily estimate the smooth component. Denoting this estimate as $\hat{x}$, the transient component is estimated as $\hat{y} = z - \hat{x}$. Therefore, the error terms satisfy $e_{\text{smooth}} = x - \hat{x} = z - y - (z - \hat{y}) = \hat{y} - y = -e_{\text{transient}}$. That is, the estimates of the two components have the same MSE. For this reason, we report a single MSE, in the following.

In the estimation procedure, because the GPs we use have zero-mean, a trick we found useful for envelope estimation was to bias the signal so that it is non-negative (non-positive) prior to upper (lower) envelope estimation. After estimating the envelope of interest, we add back the bias. This bias can be a DC bias, or a coarse lower/upper envelope obtained by using a very high value of $\sigma$ in \eqref{eqn:basicform}.
A summary of this slightly modified scheme is provided in Algorithm~\ref{algo:smooth2}.

\begin{algorithm}\caption{Smooth Component Estimation with Debiasing}\label{algo:smooth2}
\begin{algorithmic}[1]
\Require Input signal $y$
\State Set $0<\lambda_1 \ll \lambda_0$, and $0<\lambda$, $\sigma_0 \leq \sigma_1 \leq \sigma$
\State $\ell_0 \gets \hat{x}_p$ for $p = (y, \lambda, \sigma, -\infty, y)$ \texttt{\footnotesize \%lower coarse envelope}
\State $u_0 \gets \hat{x}_p$ for $p = (y, \lambda, \sigma, y, \infty)$ \texttt{\footnotesize \%upper coarse envelope}
\State $\ell \gets u_0 + \hat{x}_p$ for $p = (y - u_0, \lambda_0, \sigma_0, 0, y - u_0)$  \texttt{\footnotesize \%lower env.}
\State $u \gets \ell_0 + \hat{x}_p$ for $p = (y - \ell_0, \lambda_0, \sigma_0, y-\ell_0, 0)$  \texttt{\footnotesize \%upper env.}
\State $x^* \gets \hat{x}_p$ for $p = (y, \lambda_1, \sigma_1, \ell, u)$  \texttt{\footnotesize \%smooth component}
\end{algorithmic}
\end{algorithm}

We conducted 20 independent trials using random signals generated as described above. We iterated Alg.~\ref{algo:pseudo} long enough so that numerical convergence is achieved. This is checked by evaluating the difference between the lhs and rhs of \eqref{eqn:fixedz}. This difference should be zero in the limit, and is in fact very close to zero in our experiments -- see Fig.~\ref{fig:Exp}d. For comparison, we also produced estimates via LTI filtering, for which we used Hamming filters of varying lengths. For each trial, we computed the MSE achieved by each estimator. To better visualize, we reordered the trials in Fig.~\ref{fig:ecurve} so that the proposed algorithm's MSEs are increasing with respect to trial index.

The MSEs of the proposed method are clearly separated from those of LTI filtering. Notice also that no one LTI filter is the best -- for each filter, there is a trial for which the LTI filter performed better than the other LTI filters. These indicate that the proposed method does offer an improvement that goes beyond LTI filtering, even if different filters were used, since we expect performance variation of LTI filtering to fall somewhere in the vicinity of the four LTI curves.

\renewcommand{\sc}{0.6}
\begin{figure}
\centering
\scriptsize
\textsf{(a) Smooth Component} \\
\includegraphics[scale = \sc]{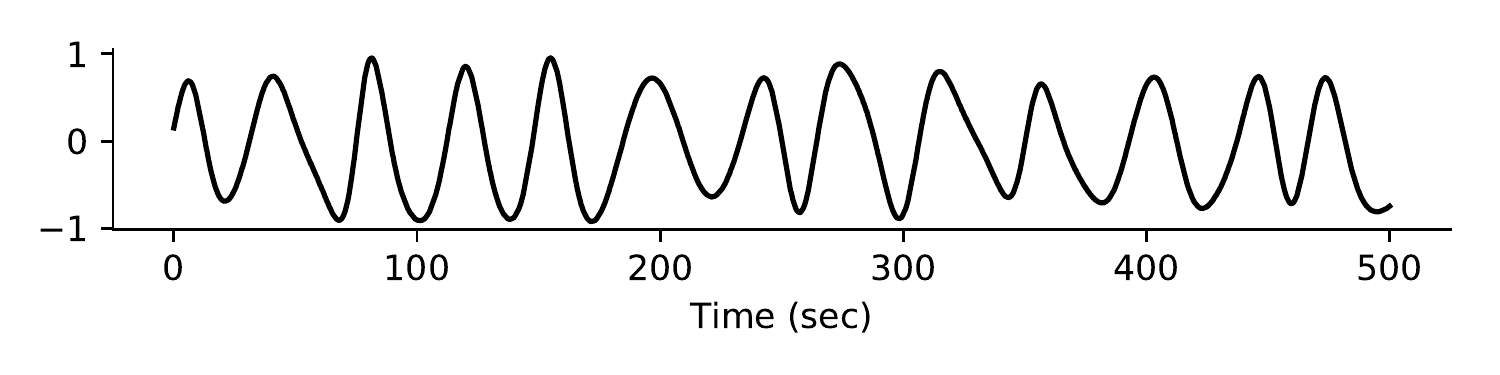}\\
\begin{minipage}{0.24\textwidth}
\centering
\textsf{(b) Transient Component} \\
\includegraphics[scale = \sc]{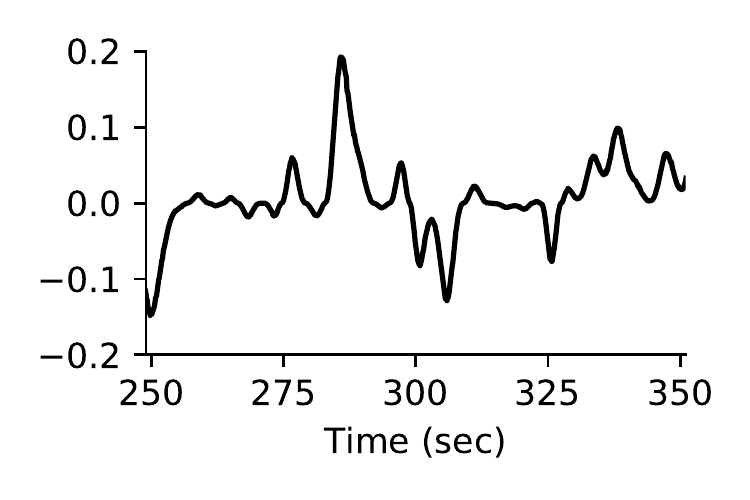}\\
\end{minipage}
\begin{minipage}{0.24\textwidth}
\centering
\textsf{(c) Observation} \\
\includegraphics[scale = \sc]{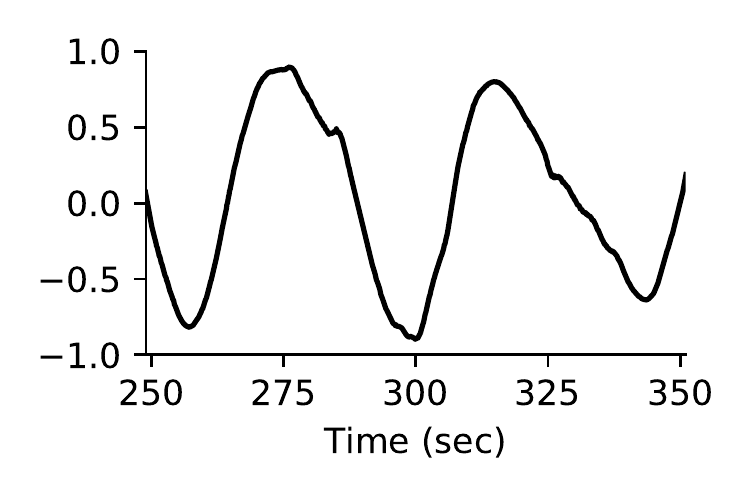}\\
\end{minipage}
\textsf{(d) Convergence Plot (lhs - rhs of \eqref{eqn:fixedz})} \\
\includegraphics[scale = \sc]{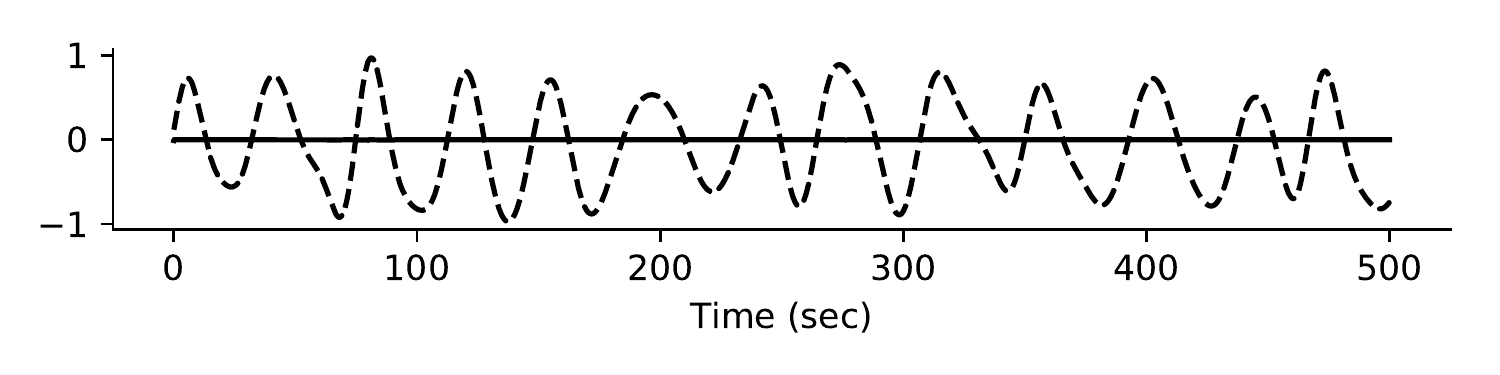}

\caption{Signals used in the one of the experiments. (a) Smooth component, (b) transient component (excerpt), (c) observation (excerpt). (d) shows the smooth part reconstructed by the method (dashed line), as well as the difference between the lhs and rhs of \eqref{eqn:fixedz}. \label{fig:Exp}}
\end{figure}

\begin{figure}
\centering
\scriptsize
\textsf{Mean Squared Error wrt Trials} \\
\includegraphics[scale = \sc]{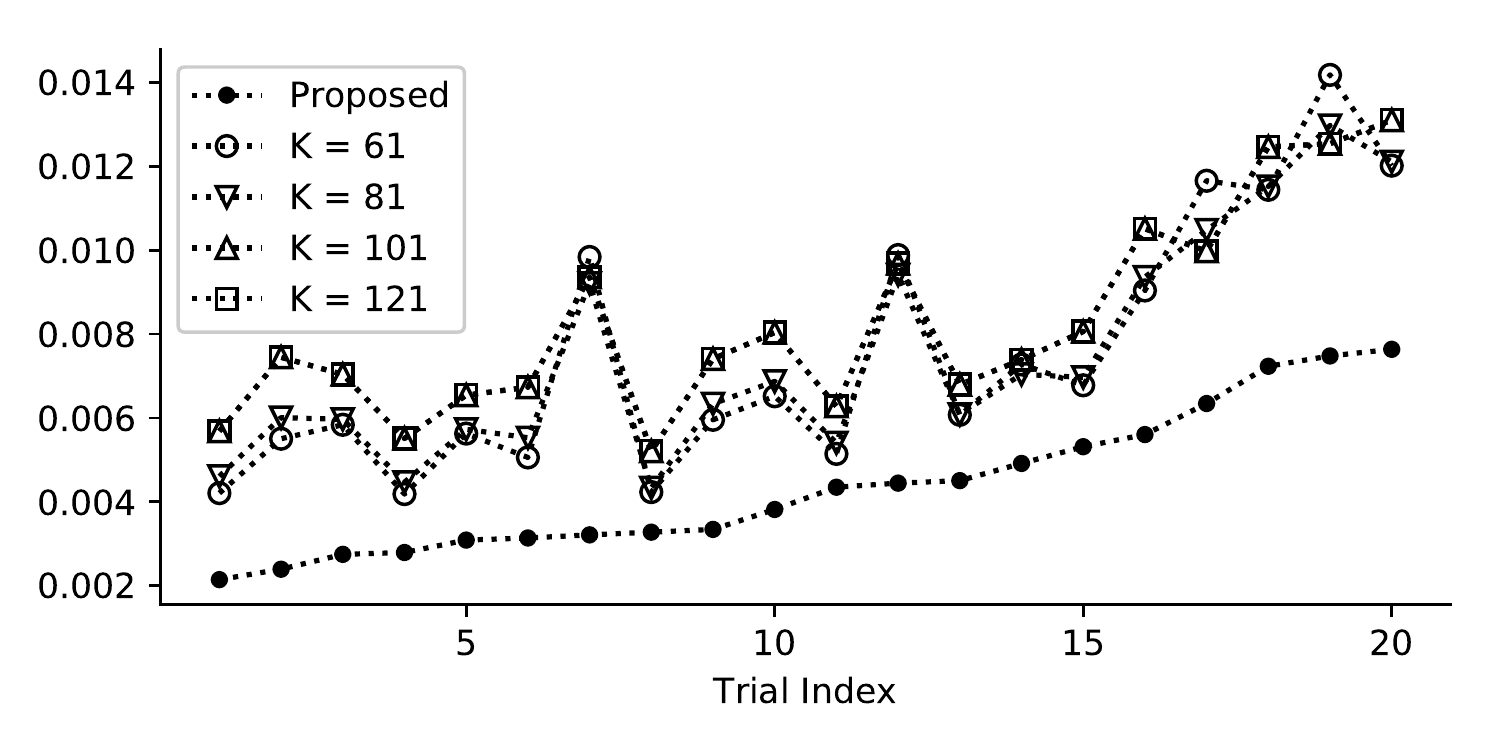}

\caption{MSEs for the 20 trials for the proposed algorithm and LTI filtering. For LTI filtering, we use Hamming filters of varying lengths. For ease of inspection, the trials are ordered such that the MSEs of the proposed algorithm increase with respect to the trial index. \label{fig:ecurve}}
\end{figure}

\subsection{Software}

Python code associated with the manuscript is available at:\\  \texttt{\footnotesize https://github.com/ilkerbayram/DualGP}

\end{document}